\documentclass[12pt]{article}
\usepackage{graphicx}
\def\gsim{\;\lower4pt\hbox{${\buildrel\displaystyle >\over\sim}$}\,}
\def\lsim{\;\lower4pt\hbox{${\buildrel\displaystyle <\over\sim}$}\,}
\def\be{\begin{equation}}
\def\ee{\end{equation}}
\def\bea{\begin{eqnarray}}
\def\eea{\end{eqnarray}}
\def\pd{\partial} 
\def\div{{\rm div}}
\def\rst{r$_\star$\,}
\def\Mst{M$_\star$\,}
\def\Bst{B$_\star$\,}
\def\dMt{$\dot {\rm M}$}

\def\Ms{M_\star}

\def\ts{t_\star}
\def\us{u_\star}
\def\rhos{\rho_\star}
\def\cF{{\cal F}}
\def\cQ{{\cal Q}}
\begin{document}
\thispagestyle{empty}
\begin{center}
\vskip 20pt
{\large Dynamics of the Flows \\
Accreting onto a Magnetized Neutron Star}
\vskip 20pt
A.M.Bykov and A.M.Krassilchtchikov \\
{\it A.F.Ioffe Institute for Physics and Technology \\
 26 Polytechnicheskaia, 194021, St.Petersburg, Russia}
\end{center}
\vskip 20pt
\begin{abstract}
Non-stationary column accretion onto a surface of a magnetized
neutron star is studied with a numerical code based on modified
first-order Godunov method with splitting. 
Formation and evolution of shocks in the column is modeled
for accretion rates ranging from 10$^{15}$ g s$^{-1}$ to 10$^{16}$
g s$^{-1}$ and surface magnetic fields ranging from
5$\cdot$10$^{11}$ G to 10$^{13}$ G.
Non-stationary solutions with plasma deceleration at
collisionless oscillating shocks are found. 
The kinetic energy of the accreting flow efficiently transforms 
into a cyclotron radiation field. 
Collisionless stopping of the
flow allows a substantial part of accreting CNO nuclei to avoid
spallation and reach the surface. The nuclei survival fraction
depends on the surface magnetic field, being higher at lower
magnetic fields.
\end{abstract}
\vskip 12pt
\noindent
{\it keywords:} accretion, neutron stars, hydrodynamics
\vskip 300pt \noindent e-mails: {\bf byk@astro.ioffe.ru} and
{\bf kra@astro.ioffe.ru} \\
Published in Astronomy Letters, vol. 30, p. 309-318 (2004)
\newpage
\setcounter{page}{2}
%

\section{Introduction}
\label{sect-intro}



Accretion onto compact objects has been being considered an efficient source of hard
emission for about forty years already (see the pioneer papers of Zel'dovich (1964),
Salpeter (1964), Pringle and Rees (1972), and Shakura and Sunyaev (1973)).


A number of successful analytical, semi-analytical, and numerical
models of accretion onto neutron stars (NSs) and black holes has
been developed by now [e.g. see the book of Frank, King, and Raine
(2002) and references therein], but several important problems already
formulated in the very earliest papers have no clear answers as
yet. One of the fundamental points is whether the matter accretes as a gas of interacting
particles ("hydrodynamic regime") or as a number of separate non-interacting particles
("freefall regime"). Zel'dovich (1967) and Zel'dovich and Shakura (1969) considered
both cases and showed that the spectrum emitted near the surface of a NS critically
depends upon the particular regime under consideration.
Up to now accretion models exploit only indirect qualitative
arguments to give prove to a particular regime in the hope that a
realistic model supported by observational evidences may a
posteriori justify their choice at least for a certain range of
global parametres of an accretion system.
Another important point is whether and when collisionless shocks (CSs) appear
and evolve in an accreting flow to decelerate the matter on its way to
the surface.

Bisnovatyi-Kogan and Fridman (1970) pointed out that a CS can appear in a flow
accreting onto a NS if the star possesses a dipole magnetic field B $\sim$ 10$^8$ G.

A shock that decelerates the accreting flow in a binary system
plays a key role in the model of Davidson and Ostriker (1973).
Shapiro and Salpeter (1975), Basko and Sunyaev (1976), Langer and
Rappoport (1982), and Braun and Yahel (1984) considered accretion
onto a NS under various assumptions on the star's magnetic field
strength and found {\it stationary} solutions for similar sets of
hydrodynamic equations used to describe the accretion flow. The
models of Shapiro and Salpeter (1975) and Langer and Rappoport
(1982) postulated the existence of a {\it stationary
collisionless} shock at a certain height over the star's surface,
the height being a model parameter.  Basko and Sunyaev (1976)
constructed an accretion model with a {\it radiative} shock in the
star's atmosphere. Braun and Yahel (1984) claimed that a {\it
stationary collisionless} shock can exist over the surface of a
magnetized NS only when the accretion rate is low enough (namely,
when the rate does not exceed a few percent of the Eddington
value).

Detailed two-dimensional non-stationary {\it radiation-dominated
super-Eddington} models of accretion onto a magnetized NS have been developed by
Arons, Klein, and collaborators (see Klein and Arons (1989), Klein et al. (1996)
and references therein).  The authors of the models showed that if the accretion
rate is much higher than the Eddington one, {\it non-stationary radiation-dominated}
shocks come out and evolve in the accretion column. An essential feature of these models
is that the existence of a shock in the column is not postulated a priori, being a result of
evolution of the accretion system.


In this paper a numerical model of {\it sub-Eddington}
one-dimensional non-stationary hydrodynamic (in the
above-mentioned sense) accretion onto a magnetized NS is
presented. The main feature of the model is a Godunov-type
numerical approach that allows to describe discontinuous flows,
and in particular, dynamics and evolution of shocks.

The model consistently takes into account kinetics of electron-ion flows in a
strong external magnetic field. The field in the accretion column is fixed, but it plays
a key role in various processes of interaction of matter and radiation that
govern the flow evolution.


In section \ref{sect-hydro} a hydrodynamic model of an accretion flow is presented.
Based on the modeled flow profiles, destruction probabilities
for accreting CNO nuclei are calculated in section \ref{sect-CNO}.
The obtained results are summarized in section \ref{sect-discuss}.


\section{Accretion Flow Hydrodynamics}
\label{sect-hydro}

\subsection{Formulation of the Problem}

Evolution of an accretion flow in a magnetic column above a
polar cap of a magnetized NS at distances upto several NS radii
from the surface is considered.

The accretion flow is assumed to be "hydrodynamic" as magnetohydrodynamic
instabilities grow very fast under typical conditions of an accretion column.
As the initial flow is highly anisotropic, multistream instabilities can have
growth rates comparable with the ion plasma frequency $\omega_{pi} \sim 1.2
\cdot 10^{12} n_{18}^{1/2}$ s$^{-1}$, where n$_{18}$ is the ion number
density measured in 10$^{18}$ cm$^{-3}$. Such growth rates are typical for
isotropisation processes in CSs. Note that the cyclotron frequencies are much
higher at considered magnetic fields. A typical propagation velocity of
MHD-waves is V$_{\hbox{Alfv\'en}}\approx c\,(1-\alpha)$, where
$\alpha$=10$^{-11}\cdot$ n$_{18}\cdot$ B$_{12}^{-2} \ll$ 1. This scaling formula
can be derived from general relations mentioned, e.g., in the book of Libermann
and Velikovich (1986).  Microscopic modeling of instabilities at NS magnetic
fields is not yet feasible, and thus it is assumed that instabilities grow with
increments of order $\omega_{pi}$.

It is assumed that electrons and ions move in the accretion flow
with the same mean (flow) velocity but have different temperatures.

\begin{figure*}
\centering
\includegraphics[height=20cm]{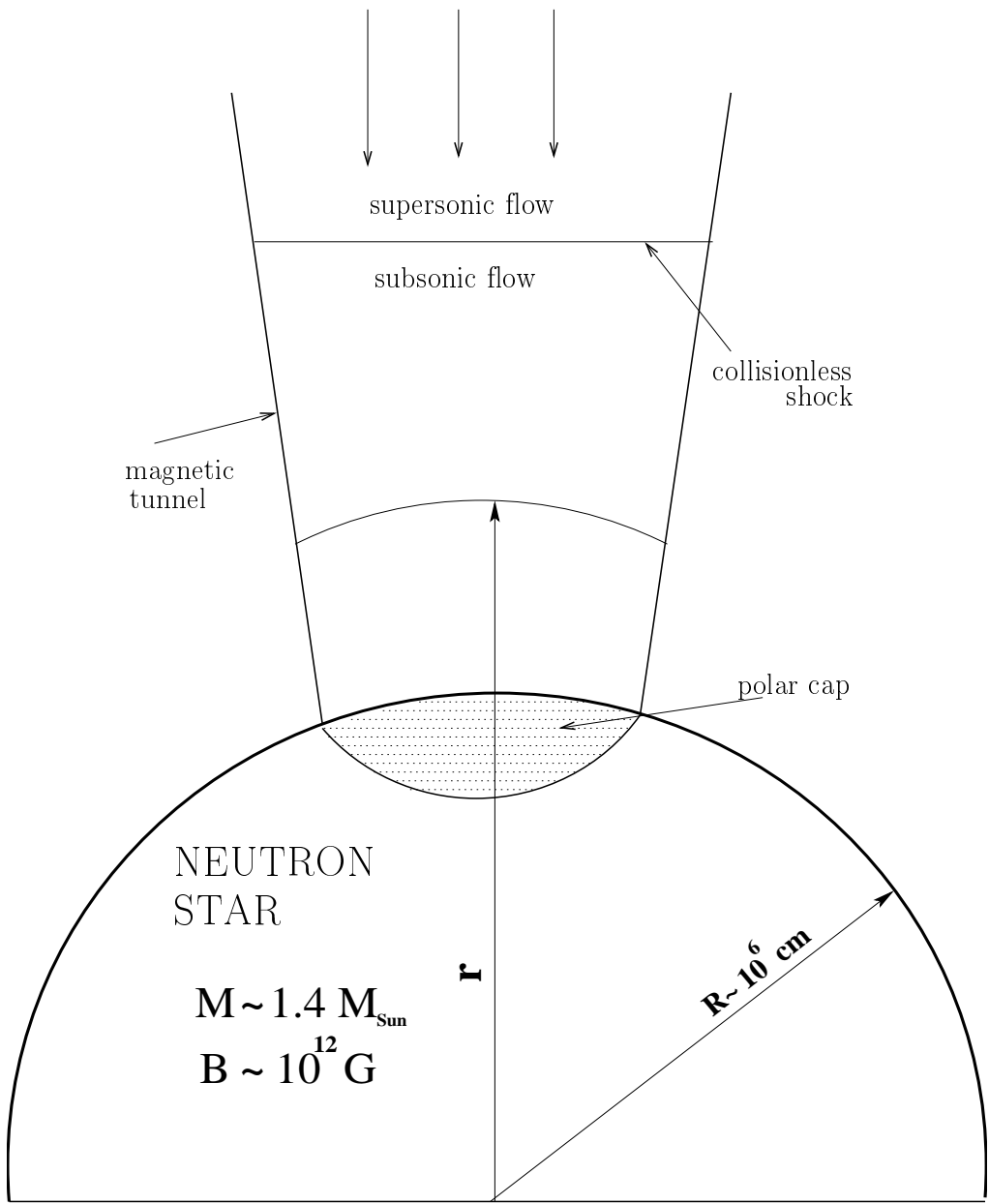}
\caption{Accretion column geometry}
\label{fig1}
\end{figure*}

The NS is assumed to possess a constant dipole magnetic field on the considered
timescales. The accretion column configuration is shown in Fig.\ref{fig1}.

\subsection{Main Parametres and Equations of the Model}

The main parametres of the model are the mass (\Mst) and the radius (\rst)
of a neutron star, as well as the magnetic field strength at its magnetic pole (\Bst)
and the accretion rate per unit area of the base of the accretion column (\dMt/A$_0$).

A basic system of hydrodynamic equations that describe flow evolution can be written as
\bea
&& {\pd \rho \over \pd t} + \div(\rho {\bf u}) = 0    \label{st1}
\\
&& {\pd (\rho u_{\alpha}) \over \pd t} + {\pd p \over \pd x_{\alpha}} +
{\pd \over \pd x_{\beta}}(\rho u_{\alpha} u_{\beta}) = \cF_{\alpha}
\\
&& {\pd \over \pd t}\left[\rho_s (E_s+{u^2 \over 2}) \right] +
    \div[\rho_s {\bf u}(E_s+{u^2 \over 2}) + p_s {\bf u}] = \cQ_s \label{st3}
\eea
where $\rho=\rho_e+\rho_i$, $p=p_i+p_e$,
$\cF$ and $\cQ_{s}$ denote sources of momentum and energy, while
$s=i,e$ denotes the sort of particles.

The system should be completed with equations of state for each sort of particles.
The equation of state for an ideal gas is used:
$E_s = p_s / [\rho_s (\gamma_s-1)]$.
For $x_s = k_B T_s/(m_s c^2) \ll 1$
the adiabatic index $\gamma_s$ may be written as
$\gamma_s \approx \gamma_{0s} (1 - x_i)$ (de Groot et al. (1980)),
where $\gamma_{0i} = $ 5/3 is a usual non-relativistic value for particles with three
degrees of freedom, while $\gamma_{0e} = $ 3, as in strong fields under consideration
the electrons are quasi-one-dimensional.

The momentum distribution of the electrons is one-dimensional, because the
typical relaxation time of electron Landau levels is about
10$^{-15}$B$^{-2}_{12}$ s, where B$_{12}$=B/10$^{12}$ G (see e.g. Bussard
1980); this time is the shortest one in the system after the cyclotron time. The
radiative decay time of the ion Landau levels is about
5$\cdot$10$^{-9}$B$^{-2}_{12}$ s that is much longer than the time of
collisionless relaxation of the ions $\propto \omega_{pi}^{-1}$. Thus, the ions
occupy highly excited Landau levels and may be considered three-dimensional and
treated quasi-classically.

{\it One-dimensional} motion of accreting plasma along {\it dipole}
lines of the star's strong magnetic field is considered. Within such a geometry the system
(\ref{st1})-(\ref{st3}) may be rewritten as
\bea
&& r^3 {\pd \rho \over \pd t} + {\pd \over \pd r} (r^3 \rho u) = 0
\label{eq1}
\\
&& r^3 {\pd (\rho u) \over \pd t} + {\pd \over \pd r} [r^3(p+\rho u^2)] =
    r^3 \cF + 3r^2 p
\\
&& r^3 {\pd \over \pd t}\left[\rho_s (E_s+{u^2 \over 2}) \right] +
{\pd \over \pd r}(r^3[\rho_s u(E_s+{u^2 \over 2}) + p_s u]) = r^3 \cQ_s.
\label{eq3}
\eea
%


The system (\ref{eq1})-(\ref{eq3}) should be completed by initial and boundary
conditions. An accretion column filled with cold free-falling gas is considered as
an initial condition for the simulations. The boundary condition in the upper part
of the column is an inflow of a cold supersonic stream, while the boundary condition
at the star's surface is the absence of any flows into the star.

\subsection{Physical Processes in an Accreting Flow}

In this section the processes
contributing to $\cF$ and $\cQ_s$ in the system (\ref{eq1})-(\ref{eq3}) are described.

\vspace*{5mm}

Since a single-velocity flow is considered, the forces acting on ions and electrons
are all summed in a single force term:
$$\cF=F^i+F^e,\ F^i=F^i_{grav}-F_{atm},\ F^e=F^e_{grav}-F_{nonres}-F_{res},$$
where
$F^i_{grav}$ and $F_{atm}$ denote gravity and viscous friction force (that is
substantial only in the thin atmosphere of the star), while $F^e_{grav}$, $F_{nonres}$,
and $F_{res}$ denote gravity, non-resonant, and resonant radiative pressure, respectively.

The force of gravitation acting on a unit volume is
\be
F_{grav}=F^e_{grav}+F^i_{grav}=(n m_i + Z n m_e){G \Ms \over r^2},
\ee
where $n$ -- is the ion number density. \\

To calculate the non-resonant radiative pressure the following formula
from the book of Zheleznyakov (1996) is adopted:

\be
F_{nonres}=n_e {\sigma_T \over c}{\sigma_{ST} T_\gamma^4 \over 1+\tau_T},
\ee
where $n_e$ is the electron number density, $\sigma_{ST}$ is the Stefan-Boltzmann constant,
$T_\gamma$ is the local temperature of radiation field, and
$\tau_T$ is the non-resonant optical depth. \\

An equation of radiation transfer in the cyclotron line is numerically integrated
to calculate the resonant radiative pressure force $F_{res} = dU_{phot} / dr$
and obtain $U_{phot}$ -- the energy density of the photon field.

As the accretion column is optically thick in the cyclotron line, the transfer
equation may be written as a diffusion equation:
\be
{\bf \nabla \cdot {\bf J_{phot}} } = S_{phot} + {1 \over 3} {\bf u \cdot \nabla} U_{phot}, \label{cy}
\ee
where $S_{phot}$ denotes the sources of cyclotron photons,
and ${\bf J_{phot}}= {4\over 3} {\bf u} \, U_{phot} - \kappa {\bf \nabla} U_{phot}$ is
the diffusive flux of photons in the cyclotron line.
As diffusion of cyclotron photons across the magnetic filed lines is substantially hampered
(see e.g. Arons, Klein, and Lea 1987), only the parallel component of (\ref{cy}) is considered,
that may be written as
\be
{1\over r^3}{\pd \over \pd r}\left\{ r^3 \left[{4\over 3} U_{phot} \, u -
\kappa_{||} {1\over r^3}{\pd \over \pd r}\left(r^3 U_{phot}\right)\right]\right\} =
S_{phot} + {1\over 3} u {1\over r^3}{\pd \over \pd r}\left(r^3 U_{phot}\right),
\ee
where $\kappa_{||}$ is the diffusion coefficient parallel to the magnetic field.
The equation is integrated by a shooting method with the following boundary conditions:
cyclotron photons freely leave the upper border of the column, while their number density
on the star's surface is determined by a blackbody spectrum of temperature T$_{eff}$. The
surface temperature T$_{eff}$ varies with the part of flow energy released at the surface. \\

To calculate the friction force acting on the flow in the star's atmosphere
a standard expression for Coulomb stopping in a dense environment is used:

\be
F_{atm}={4\pi n_a n_i e^4 Z^2 \ln\Lambda \over m_e u^2},
\ee

where $\Lambda$ is a Coulomb logarithm, $n_a$ is the electron number density in
the atmosphere, and $u$ is the flow velocity. A similar expression was used in
the paper of Bildsten et al (1992) devoted to collisional destruction of CNO
nuclei in a NS atmosphere.

\vspace*{5mm}

The forces acting on the flow perform a certain work on it, and this work is
effectively redistributed between the ions and the electrons.

Let an external force $F^i$ act on the ions and an external force $F^e$ act on
the electrons, then it can derived from the local electroneutrality (if the
frequencies of variations of the external forces are much lower that the plasma
frequencies) that the fluxes of both types of particles are equal, i.e.

$$(F^e-eE){n_e \over m_e \nu_{ei}}=(F^i+ZeE){n_i \over m_i \nu_{ei}},$$

where $E$ is the ambipolar electric field, and $\nu_{ei}$ is the effective electron-ion
relaxation frequency.
Since $n_e=Zn_i$,
$$eE = F^e {1 \over \xi+1} - F^i {\xi \over Z(\xi+1)},$$
where $\xi=m_e/m_i$, and the resulting effective force acting of the ions is
$$F^i_{\rm eff}=(F^i+F^e Z) {1 \over \xi+1},$$
while the resulting effective force acting on the electrons is
$$F^e_{\rm eff}=(F^i+F^e Z) {\xi \over Z(\xi+1)}.$$

\vspace*{5mm}

\noindent The temperature of the ions changes in the following processes: \\

\noindent (i)\hspace*{0.27cm} small-angle e-i Coulomb scatterings ("heat exchange"): H$_{ie}$ \\
(ii) \ collisional excitation of electron Landau levels: $Q_{cyc}$ \\
(iii) collisional Coulomb relaxation in the star's atmosphere: $Q_{relax}$ \\
(iv) the work of the effective force: $F^i_{\rm eff} u$ \\

The model also accounts for the cyclotron cooling of the ions. This effect is
noticeable if the star's dipole magnetic field exceeds 5$\cdot$10$^{11}$ G and
the column is transparent for the photons of the ion cyclotron fundamental line.
\\

\noindent The temperature of the electrons changes in the following processes: \\

\noindent (i)\hspace*{0.27cm} small-angle e-i Coulomb scatterings ("heat exchange"): H$_{ei}$ \\
(ii) \ Bremsstrahlung cooling in e-i and e-e collisions: $Br_{ei}$ ╦ $Br_{ee}$ \\
(iii) collisional excitation of Landau levels in e-i and e-e collisions: $Cyc_{ei}$ ╦ $Cyc_{ee}$ \\
(iv) Compton processes: $Q_{compt}$  \\
(v) the work of the effective force: $F^e_{\rm eff} u$ \\

\noindent For a detailed description of the terms above, see the Appendix.

\subsection{The Numerical Approach}

A multicomponent accretion flow onto the surface of a NS can have discontinuities,
and in particular, shocks. That is why the chosen numerical approach is based on
a well-known Godunov method (see, e.g. Godunov 1959; 1964).

The traditional form of Godunov method is only applicable to one-component
systems\footnote{
For a review of modern multi-component methods see, e.g., Zabrodin and Prokopov (1988).}
without source terms. To account for the source terms in
(\ref{eq1})-(\ref{eq3}) a numerical scheme based on the work of
LeVeque (1997) is exploited. The system (\ref{eq1})-(\ref{eq3}) is split into two parts: one
of them consists of flux-conserving terms and is integrated with a modified
Godunov method, while the other describes the sources of energy and momentum and
is integrated with a Gear method.

Such a way of solving the system is due to its complex structure with two sorts
of particles non-linearly interacting with each other,
with external magnetic and gravitational fields as well as with the radiation field.

To make the system (\ref{eq1})-(\ref{eq3}) dimensionless $\cF$ is multiplied
by $C_F = {\ts \over \rhos \us}$, while $\cQ_s$ is multiplied by $C_Q = {\ts \over \rhos \us^2}$,
where $\ts$, $\us$, and $\rhos$ are scales of time, velocity, and density, respectively.

Integration of equations (\ref{eq1})-(\ref{eq3}) in the accretion column is performed
in a combined way that allows to generalize the standard Godunov method for systems with
energy and momentum exchange between the components (i.e. account for source term in
(\ref{eq1})-(\ref{eq3})). The accretion column is represented by a number of spatial cells.
The system is integrated from an initial state at the moment
$t=0$ to a current state at a moment $t$ in a number of time steps $\Delta t$ each of
them consisting of the following stages: \\

(i) integration of flux-conserving terms without source terms: \\

\bea
&& r^3 {\pd \rho \over \pd t} + {\pd \over \pd r} (r^3 \rho u) = 0
\label{p11}
\\
&& r^3 {\pd (\rho u) \over \pd t} + {\pd \over \pd r} [r^3(p+\rho u^2)] = 0
\\
&& r^3 {\pd \over \pd t}\left[\rho_s (E_s+{u^2 \over 2}) \right] +
{\pd \over \pd r}(r^3[\rho_s u(E_s+{u^2 \over 2}) + p_s u]) = 0
\label{p13}
\eea

At this stage the system is simultaneously integrated in all the cells.

(ii) integration of source terms in individual cells: \\

\bea
&& {\pd (\rho_i u_i) \over \pd t} = C_F \cF_i  + {3 \over r_i} p_i
\label{p21}
\\
&& {\pd \over \pd t}\left[\rho_{s_i} (E_{s_i}+{u_i^2 \over 2}) \right] = C_Q \cQ_{s_i},
\label{p22}
\eea

where $q_i$ are cell-averaged quantities in the cell number {\it i}.

At each of the two stages the system is integrated over the same time step
determined by the Courant condition at the first stage.

The system (\ref{p21})-(\ref{p22}) is stiff and thus it is integrated with
a standard LSODE routine (Hindmarsh 1983) that implements the Gear-B method
(see, e.g. Gear 1971). It should be noted that a number of modern methods for
integration of stiff systems based on the modified Bulirsch-Stoer method
(see Press et al. 1993) require an explicit Jacobian of the integrated system to
be supplied. As the system (\ref{p21})-(\ref{p22}) has a very complex right-hand
part with non-analytical terms, the Bulirsch-Stoer methods appear too complicated
to implement.

The system (\ref{p11})-(\ref{p13}) is integrated with a "capacity-differencing"
modification of the standard first-order Godunov method suggested by LeVeque (1997).

The essence of "capacity-differencing" is the following. If a conservation law for
a physical quantity q(x,t) has a generalized form
$$ \kappa(x) {\pd q(x,t) \over \pd t} + {\pd f(q(x,t)) \over \pd x} = 0, $$
where $\kappa$(x) is a known function of the space coordinate, denoting effective
"capacity" (e.g. porosity of the medium) the traditional
Godunov value of net function q at the moment $t_0 + \Delta t$
$$ \tilde q_i = q_i - {\Delta t \over \Delta x_i}\left( F_i - F_{i+1} \right), $$
where $F_i$ is the flux of $q$ from the i{\it th} cell to the (i-1){\it th} cell
should be replaced by
$$ \tilde q_i = q_i - {\Delta t \over \kappa_i \Delta x_i}\left( F_i - F_{i+1} \right),
$$  where $\Delta x_i$ is the size of the i{\it th} cell, while $\kappa_i$ is the
cell-averaged value of $\kappa$(x). \\

\subsection{Simulation Results}

Evolution of an accreting flow has been simulated according to the scheme
described above for a number of sets of global parametres.

\begin{figure*}
\centering
\includegraphics[height=18cm]{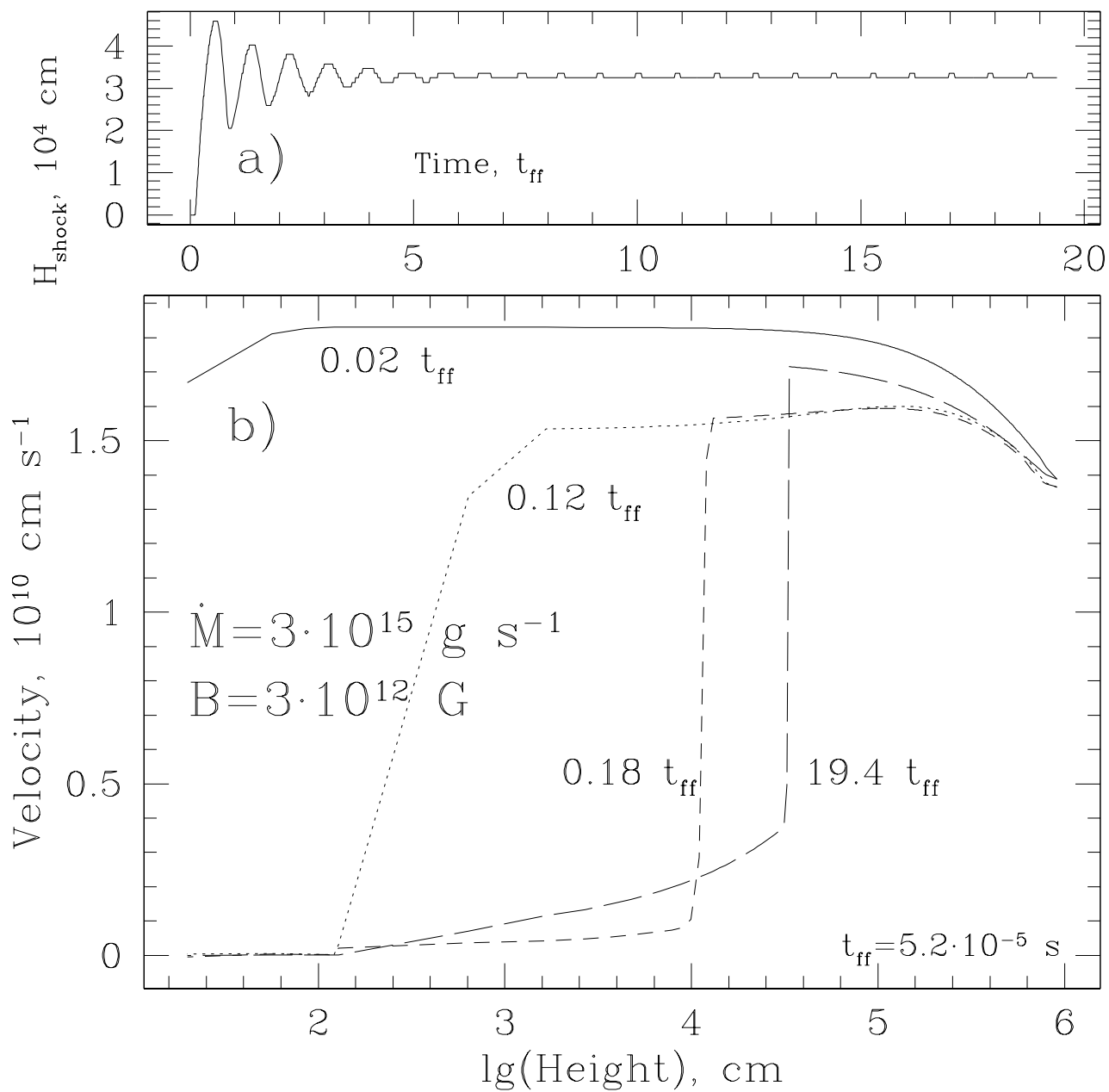}
\caption{Evolution of an accreting flow: a) shock height vs time \hspace*{9cm}
                     b) evolution of velocity profile}
\label{fig2}
\end{figure*}

Strong shocks have been found to evolve in the column on timescales
of about 10$^{-5}$~s. The shocks oscillate around their equilibrium positions
with periods of about 10$^{-5}$~s, the oscillations dumping time
being about 10$^{-3}$~s. A typical evolution of a shock and of
a flow velocity profile is presented in Fig.\ref{fig2}.

\begin{figure*}
\centering
\includegraphics[height=18cm]{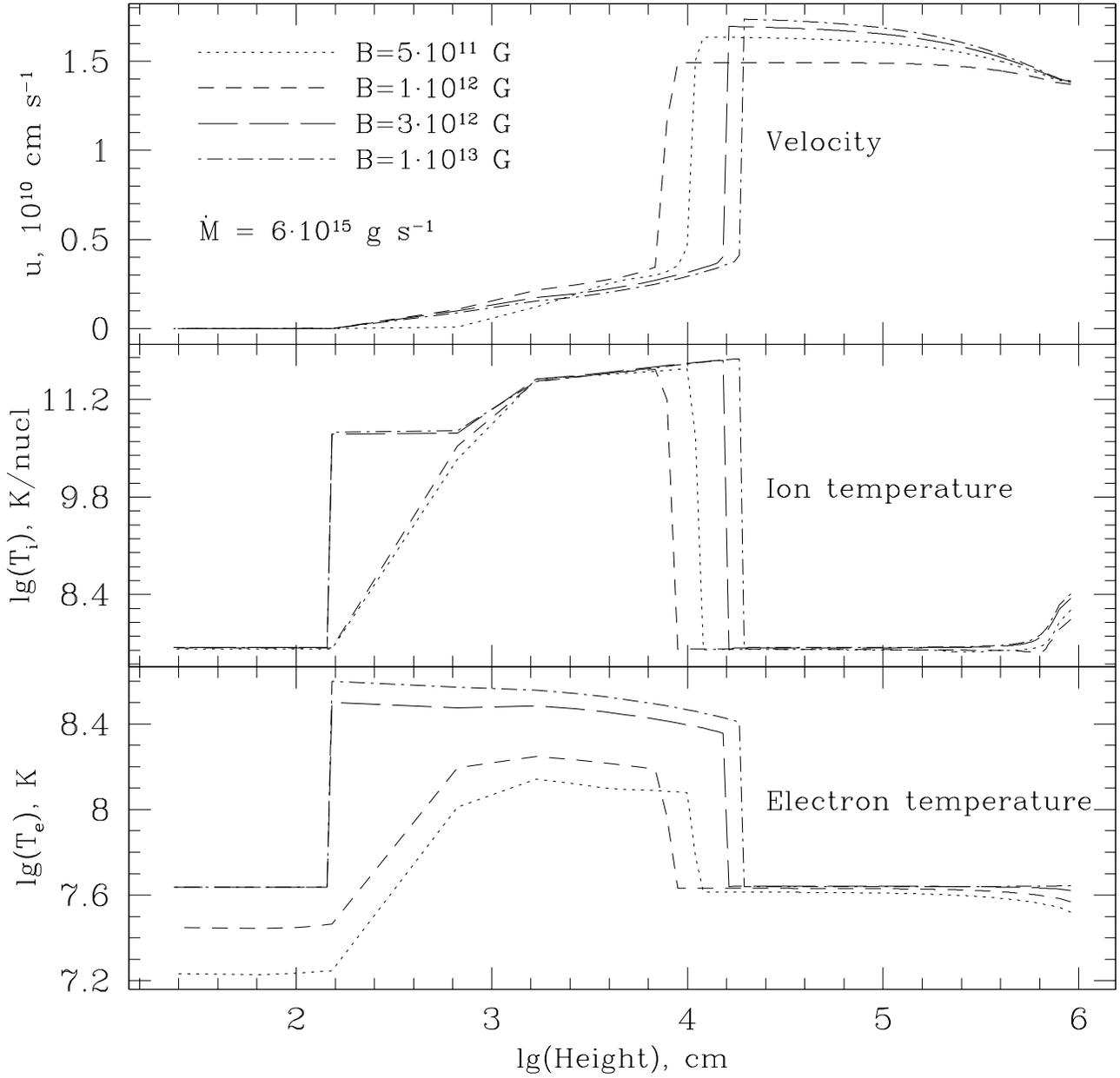}
\caption{Flow profiles at different magnetic field values}
\label{fig3}
\end{figure*}

Modeled flow profiles in Figs. \ref{fig3} and \ref{fig4} demonstrate
stable and strong shocks that decelerate and heat the accreting flow.
At such shocks the ions are heated much more than the electrons as they
contain most of the kinetic energy of the flow. However, as the hot ions move
towards the surface in the downstream zone, they pass a substantial part of their energy
to the electrons. In their turn, the electrons emit the energy as cyclotron and
Bremsstrahlung photons and pass it to non-resonant photons in Compton collisions.

In most of the modeled cases the compression ratio at the shocks slightly
exceeds 4 (the maximal value for non-relativistic single-fluid shocks) due to
weak relativistic changes in the adiabatic index of the ions heated upto a few tens MeV.

\begin{figure*}
\centering
\includegraphics[height=18cm]{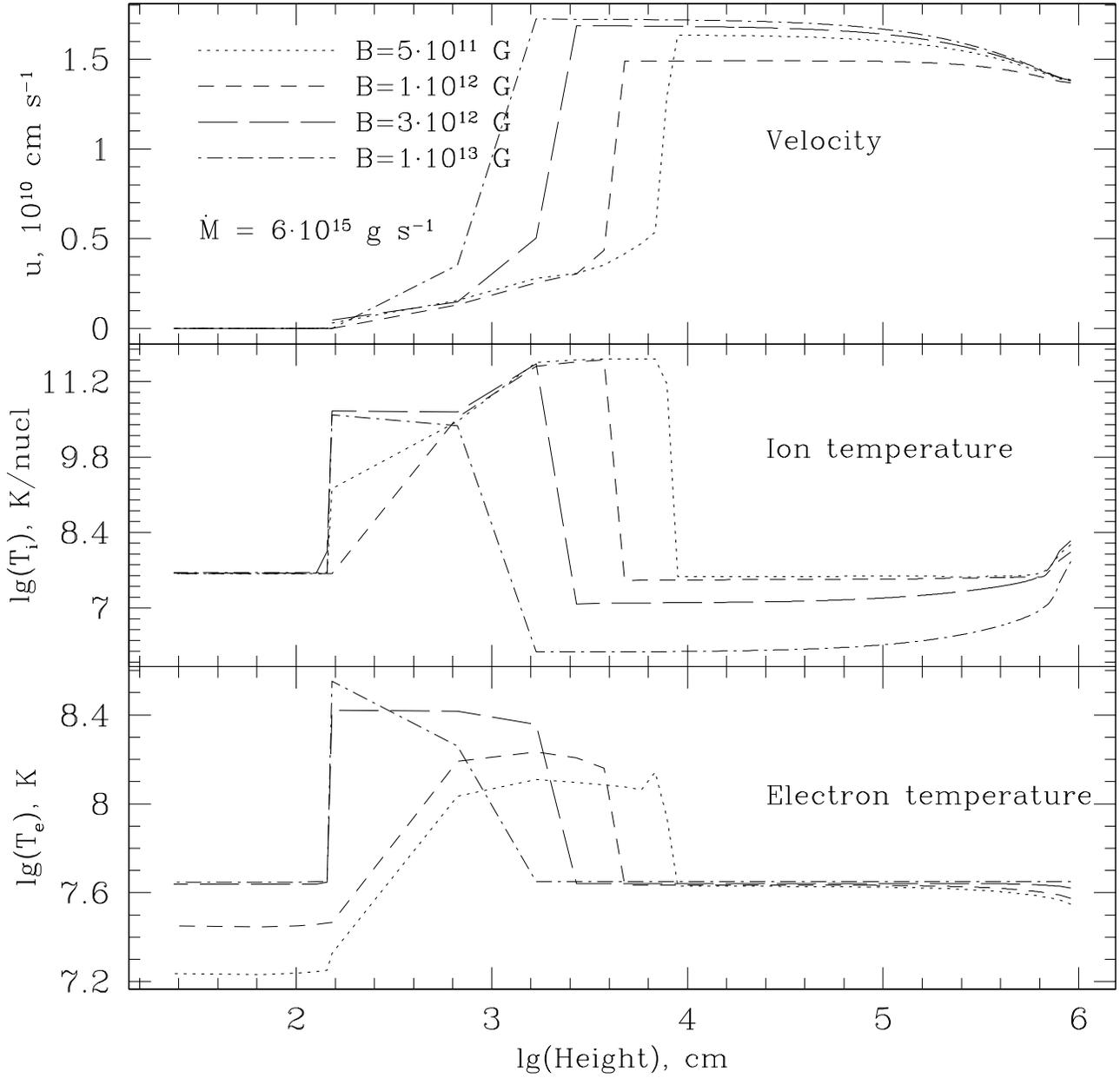}
\caption{Flow profiles at different magnetic field values when
     the column is optically thin for ion cyclotron emission}
\label{fig3a}
\end{figure*}

\begin{figure*}
\centering
\includegraphics[height=18cm]{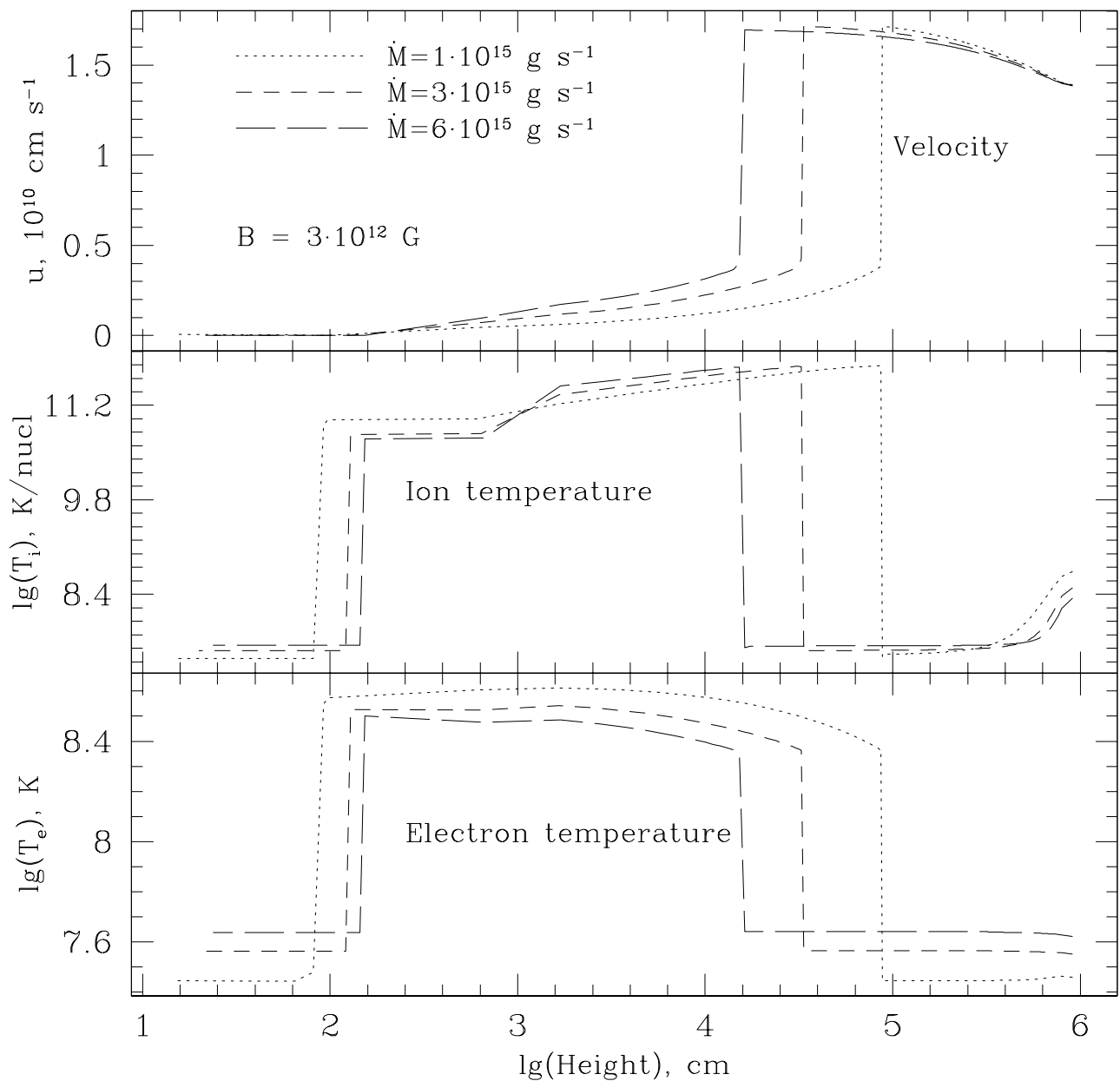}
\caption{Flow profiles at different accretion rate values}
\label{fig4}
\end{figure*}

An important property of the model is transformation of a substantial part
of the ram energy of the flow into cyclotron photons of the optically thick line.
Fig.\ref{fig6} demonstrates the dependence of the part on magnetic field strength.
The pressure of the trapped cyclotron photons noticeably affects
the stopping of the accreting flow. The cyclotron cooling of the ions is significant
at magnetic fields exceeding 5$\cdot$ 10$^{11}$ G; in this case, the particular accretion regime
significantly depends upon the detailed structure of magnetic fields in the star's
atmosphere about 10$^3$ cm from the surface. At such heights the field may be
significantly non-dipolar due to the presence of local high-multipole components.
The non-uniform structure of the field makes the column transparent for optical and X-ray
cyclotron photons emitted by ions and electrons.
Accretion regimes for such an optically thin column where the cyclotron emission of the
ions freely comes out are demonstrated in Fig.\ref{fig3a}. An account for cyclotron cooling of the ions at magnetic fields
exceeding 5$\cdot$ 10$^{11}$ G leads to a significantly lower position of the shock. In this
case the emission spectrum will most probably contain a prominent proton cyclotron line
in the optical/UV range. Due to strong deceleration and efficient cooling of the flow in
an optically thin part of the column only about half of the flow energy reaches the surface
of the star (see Fig.\ref{fig7}).

If magnetic field structures are regular at the heights of about 10$^3$ cm
from the surface, the proton cyclotron line may become optically thick.
In this case the accretion regime is analogous to that in Fig.\ref{fig3},
as the high-frequency collisionless relaxation recovers the isotropy of the
ion distribution much faster than the transverse temperature changes.

It should be also noted that the qualitative difference in flow profiles
at low and high values of magnetic field (Fig.\ref{fig3}) is due to the fact,
that at low fields the electron temperature in the downstream region is lower
and the gradient of locked cyclotron photons energy density increases.
A large gradient makes the flow to decelerate more efficiently.


\section{Destruction of CNO Nuclei \\ in an Accretion Column}
\label{sect-CNO}

An important characteristic of an accreting flow is the chemical
composition of the matter that reaches the surface of a NS. The
characteristic is particularly important for the theory of X-ray
bursts, as accreted CNO nuclei may catalyze thermonuclear burning
of hydrogen on the surface of a NS (see Lewin, van Paradijs, and
Taam 1993, Strohmayer and Bildsten 2003, and references therein).

Most of the bursts (Type I bursts) have been
observed in low-mass X-ray binary systems where the NSs
are usually thought to possess magnetic fields
not exceeding 10$^9$ G, while the model presented here
treats much stronger fields usually attributed to high-mass X-ray
binary systems. Nevertheless, it is instructive
to study the dependence of the survival fraction on 
magnetic field and mass accretion rate in the scenario of collisionless 
matter stopping at an accretion
shock wave described above.

The nuclei spallation process
leads to fast destruction of tens MeV/nucl range nuclei and thus,
to a significant suppression of gamma-ray line emission from
accreting objects
(see e.g. Aharonian and Sunyaev 1984 and Bildsten et al. 1992).
Bildsten et al. (1992) considered the destruction scenario for the
case when the accretion flow is decelerated by Coulomb collisions
in a dense atmosphere of a NS and concluded, that almost all the
CNO nuclei would be destroyed before they reach the surface. The
authors noted, however, that this conclusion may be not true in
the case when the flow decelerates in a collisionless column above
the atmosphere as it is in our model.

Indeed, in that case, the depth traversed by a nucleus as it
decelerates down to energies of about 10 MeV/nucl can be much
smaller than the depth for a pure Coulomb deceleration
down to the same energies. Having modeled flow profiles in an
accretion column, one is able to say how efficient the destruction
of CNO nuclei in an accretion flow is and where they are
destroyed.

In order to test the destruction efficiency quantitatively the
destruction probability for a carbon nucleus that accretes inside
a modeled flow was calculated (the destruction crossections of
nitrogen and oxygen are very similar to that of carbon, so they
would be destroyed in the same way).

The carbon nuclei destruction rate
[with crossections given by Read and Viola (1984)] was
numerically integrated
within the flow to obtain the nuclei survival fractions as a function of
the distance from the NS surface.

\begin{figure*}
\centering
\includegraphics[height=18cm]{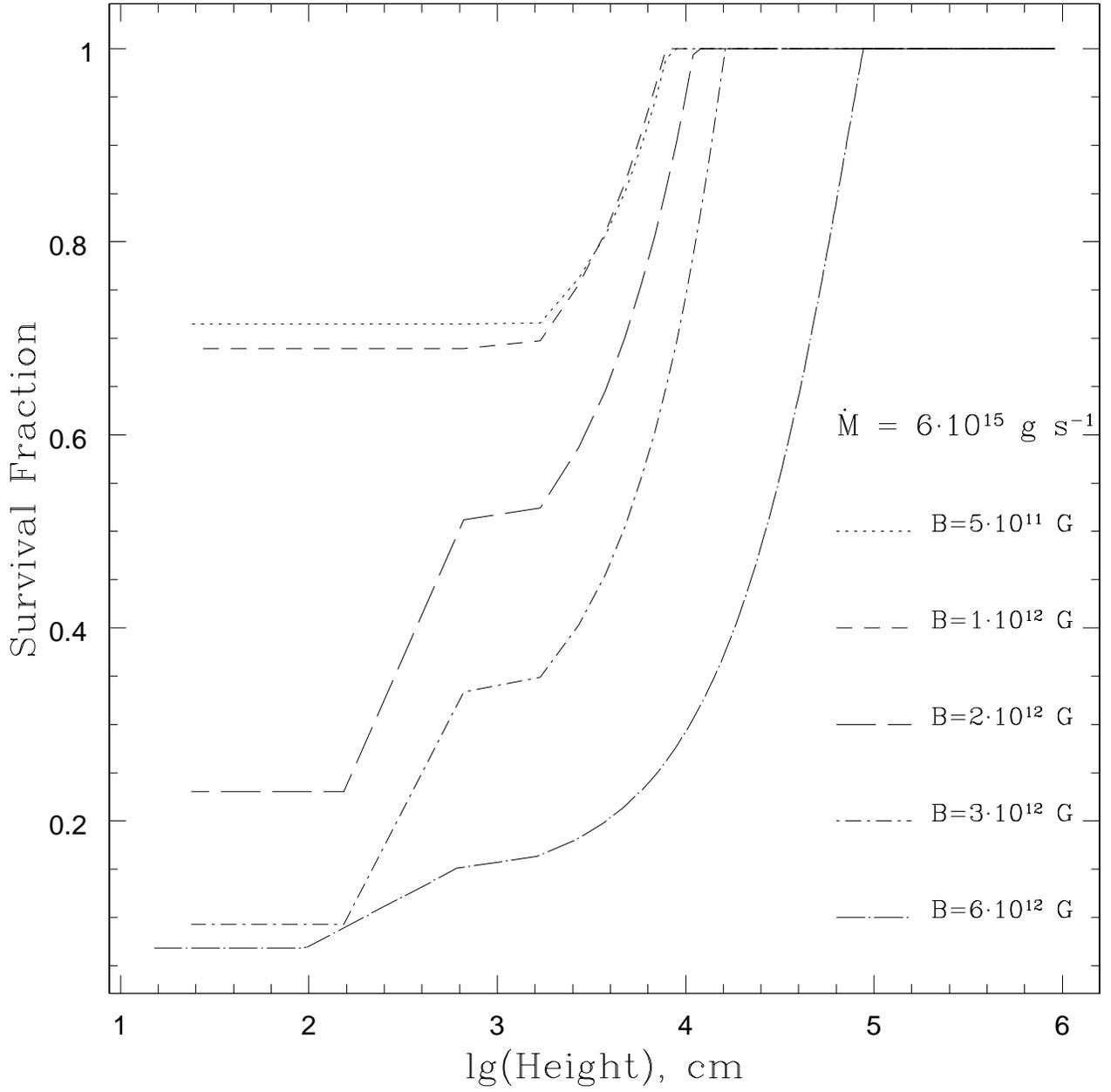}
\caption{Survival fraction of $^{12}$C nuclei
     on their way downto the surface}
\label{fig5}
\end{figure*}

\begin{figure*}
\centering
\includegraphics[height=18cm]{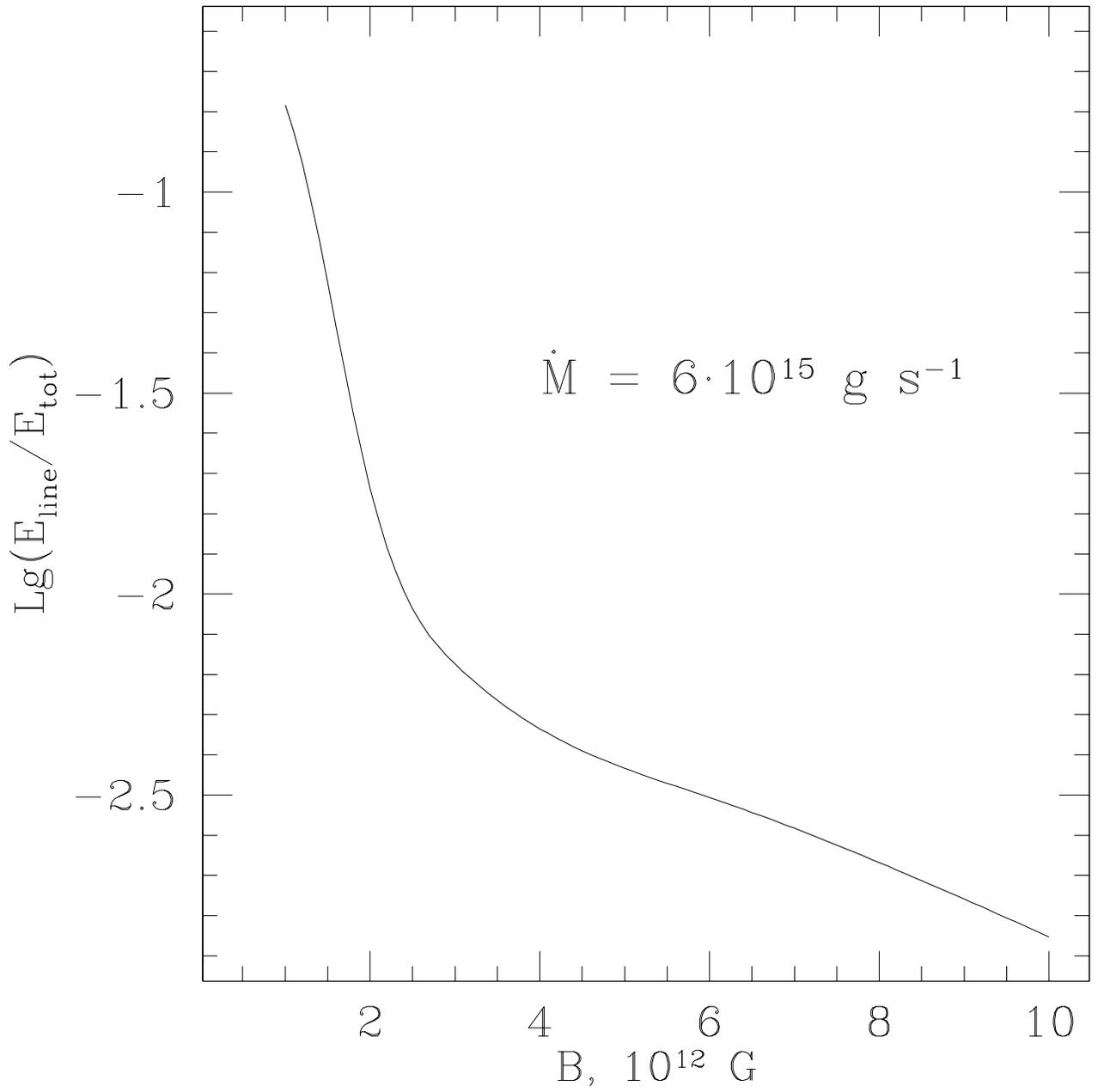}
\caption{Part of flow energy converted into cyclotron emission
at different magnetic field values}
\label{fig6}
\end{figure*}

\begin{figure*}
\centering
\includegraphics[height=18cm]{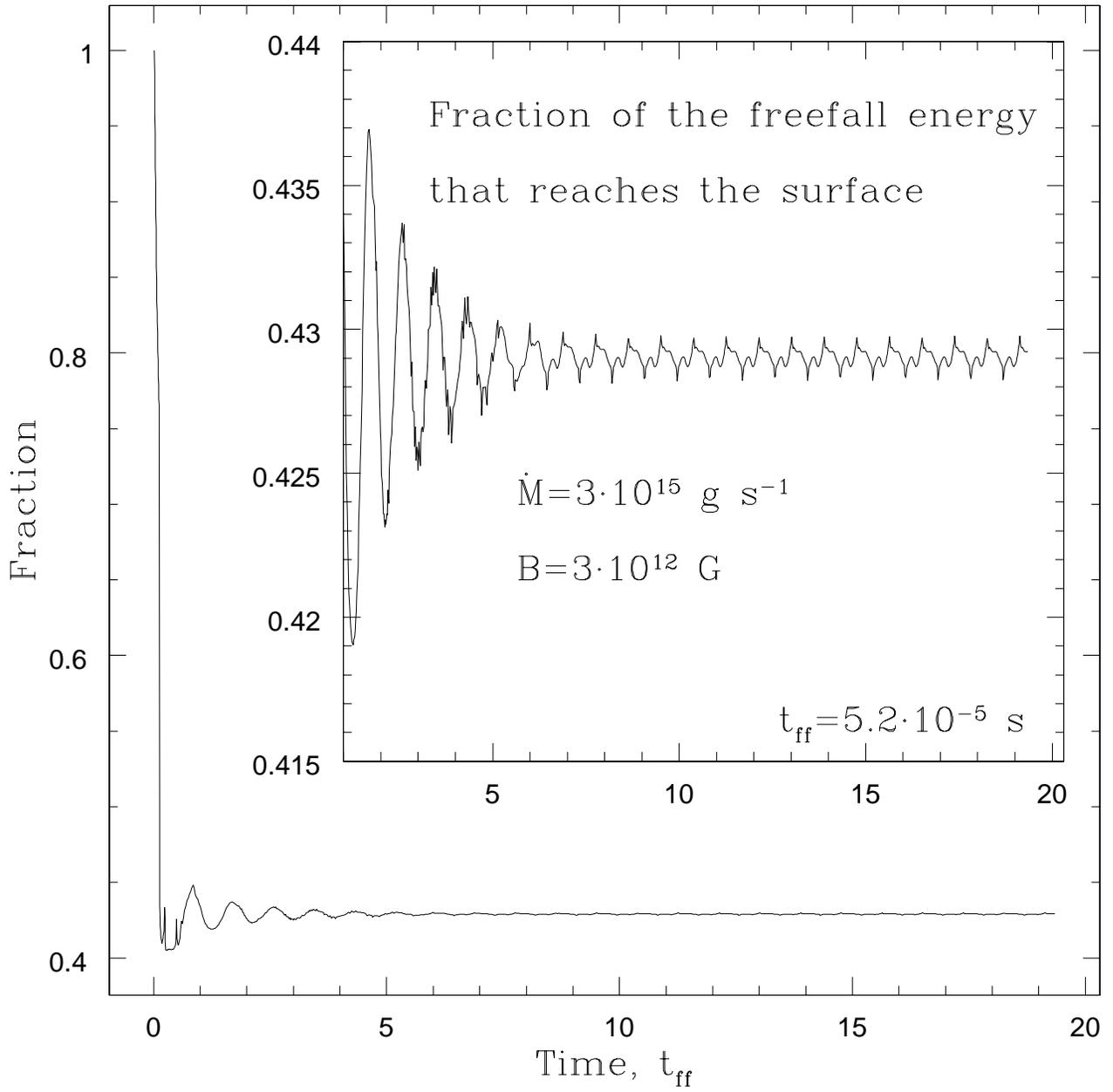}
\caption{Temporal evolution of the part of flow energy that reaches the surface}
\label{fig7}
\end{figure*}

The dependencies of survival fractions for a carbon nucleus on the distance
from the NS surface are presented in Fig.\ref{fig5} for a set of magnetic field values.
It follows that the survival fraction of C nuclei drops by
an order of magnitude for magnetic fields increasing from 5$\cdot 10^{11}$ G to
 5$\cdot 10^{12}$ G.  For magnetic fields about 5$\cdot 10^{11}$ G
 a significant fraction of the nuclei may reach the surface of the star. The survival fraction is
not too sensitive to the accretion rate in the interval studied
(from 10$^{15}$ g s$^{-1}$ to 10$^{16}$ g s$^{-1}$). The effect
may prevent Type I bursts for NSs in HMXBs with magnetic fields
$\gsim 3 \cdot 10^{12}$ G, while the bursts are allowed for lower
fields, if the hydrodynamic accretion regime considered in the
paper realizes.


\section{Summary}
\label{sect-discuss}


A numerical model of a non-stationary one-dimensional accretion
column over a polar cap of a magnetized neutron star has been
constructed. The model describes a two-fluid flow of accreting
plasma in a strong dipolar magnetic field. One of the main
features of the model is an ability to treat consistently flow
discontinuities (including shocks) with a modified Godunov method
that made it possible to study temporal evolution of shocks in the
accreting flow.

After several free-fall periods a quasi-stationary state of the column
with a stable accretion shock is usually reached. At the shock the ion temperature
jumps upto $\sim$ 10$^{11}$ K (this value weakly depends upon the magnetic field strength
and accretion rate). Depending on the magnetic field strength the electron
temperature reaches (3-5)$\times$10$^8$ K.

A part of the kinetic energy of the flow is transformed into
emission of a thick cyclotron line. The radiation pressure in the line
significantly affects the deceleration of the plasma flow. A substantial
part of the kinetic energy of the flow is emitted into the optically thin
part of the spectrum well before the flow reaches the bottom of the column.
It is often assumed that the kinetic energy of the flow and the emission
of the optically thin part of the column are transformed into blackbody
radiation in the optically thick part of the star's atmosphere.
As a significant part of the flow energy is emitted far above the atmosphere
the effective temperature of the polar cap $T_{eff}$ is reduced and
the usual relation $T_{eff} \propto {\dot M}^{1/4}$ is not true anymore, because
the part of flow energy emitted above the surface is now a complicated function of \dMt.

\vspace*{1cm}

\noindent The authors wish to thank S.P.Voskoboinikov, A.Ye.Kalina, Yu.A.Kurakin,
K.P.Levenfish, A.Yu.Potekhin, Yu.A.Uvarov, D.G.Yakovlev, as well as the participants
of V.S.Imshennik's workshop for helpful discussions and advice. We are also grateful
to the referees for constructive suggestions.

This present work was supported by the following grants:
INTAS-ESA 99-1627, RFBR 01-02-16654, RFBR 03-02-17433, RFBR 03-07-90200.


\section*{References}

\begin{list}{}{}

\item
Aharonian, F.A. and Sunyaev, R.A., 1984, MNRAS, {\bf 210}, 257

\item
Arons, J., Klein, R.I., and Lea, S.M., 1987,  ApJ, {\bf 312}, 666

\item
Basko, M.M., and Sunyaev, R.A., 1976, MNRAS, {\bf 175}, 395

\item
Berestetskii, V.B. and Landau, L.D. 1982, Quantum Electrodynamics \\
    (Betterworth-Heinemann Publishers)

\item
Bildsten, L., Salpeter, E.E., and Wasserman, I., 1992,  ApJ, {\bf 384}, 143

\item
Bisnovatyi-Kogan, G.S., and Fridman, A.M., 1970, Soviet Astronomy, {\bf 13}, 566

\item
Braun, A., and Yahel, R.Z., 1984, ApJ, {\bf 278}, 349

\item
Bussard, R.W., 1980,  ApJ, {\bf 237}, 970

\item
Davidson, K., and Ostriker, J.P., 1973, ApJ, {\bf 179}, 585

\item
Frank, J., King, A., Raine, D.J., 2002,
Accretion Power in Astrophysics (Cambridge University Press)

\item
Gear, ш.W., 1971, Numerical initial value problems in ordinary differential equations.
(Englewood Cliffs, N. J.: Prentice-Hall)

\item
Godunov, S.K., 1959, Mat.Sb., {\bf 47}, 271 (in Russian)

\item
Godunov, S.K., and Rjabenkii, V.S., 1964,
Introduction to the Theory of Difference Schemes
(Wiley-Interscience, New York)

\item
de Groot, S.R., van Leeuwen, W.A., and van Weert, Ch.G., 1980,
Relativistic Kinetic Theory: Principles and Applications
(Elsevier Science Publishers)

\item
Haug, E., 1975, Zeitschrift f\"ur Naturforschung, {\bf 30a}, 1099

\item
Hindmarsh, A.C., 1983,
ODEPACK: a systematized collection of ODE solvers
(in: Scientific Computing, eds. R.S.Stepleman et al., North-Holland Publishers,
Amsterdam, 1983, p.55)

\item
Klein, R.I., and Arons, J., 1989,
Proc. 23rd ESLAB Symp. on Two-Topics in X-Ray Astronomy, Ed. N.White,
ESA SP-296, Noordwijk, p. 89

\item
Klein, R.I., Arons, J., Garrett, J., and Hsu, J.J.-L., 1996, ApJ, {\bf 457}, L85

\item
Langer, S.H., 1981, Phys.Rev. D, {\bf 23}, no.2, 328

\item
Langer, S.H., and Rappoport S., 1982, ApJ, {\bf 257}, 733

\item
LeVeque, R.J., 1997, J. of Computational Phys., {\bf 131}, 327

\item
Lewin, W.H.G, van Paradijs, J., and Taam, R.E., 1993, Space Sci. Rev., {\bf 62:3/4}, 223

\item
Libermann, M.A. and Velikovich, A.L., 1986,
Physics of Shock Waves in Gases and Plasmas, (Springer Verlag)

\item
Press, W.H., Flannery, B.P., Teukolsky, S.A., and Vetterling, W.T., 1993,
Numerical Recipes in FORTRAN 77: The Art of Scientific Computing,
(Cambridge University Press)

\item
Pringle, J.E., and Rees, M.J., 1972, A\&A, {\bf 21}, 1

\item
Read, S.M., and Viola, V.E., 1984, Atomic Data and Nuclear Data Tables, {\bf 31}, no. 3, 359

\item
Salpeter, E.E., 1964, ApJ, {\bf 140}, 796

\item
Shakura, N.I., Sunyaev, R.A., 1973, A\&A, {\bf 24}, 337

\item
Shapiro, S.L., and Salpeter, E.E., 1975, ApJ, {\bf 198}, 671

\item
Strohmayer, T.E., and Bildsten, L., 2003, astro-ph/0301544

\item
Zabrodin, A.V., and Prokopov, G.P., 1998, Math. Modeling of
Physical Processes, series, iss.~3, p.~3 [in Russian]

\item
Zel'dovich, Ya.B., 1964, Sov.Phys.Dokl., {\bf 9}, 246

\item
Zel'dovich, Ya.B., 1967, Transactions of XIII IAU meeting, Prague

\item
Zel'dovich, Ya.B., and Shakura, N.I., 1969, Soviet Astronomy, {\bf 13}, 175

\item
Zheleznyakov, V.V., 1996, Radiation in astrophysical plasmas
(Dordrecht Kluwer Academic Publishers)

\end{list}


\section*{Appendix: Energy Exchange Rates}
\label{sect-appB}

\begin{enumerate}

\item The rates of energy exchange due to small-angle Coulomb collisions
$H_{ei}$ (and $H_{ie}=-H_{ei}$) were adopted from the work of Langer and Rappoport (1982):
\be
H_{ei}= 2\sqrt{2 \over \pi} r_0 n_e n \xi {T_i-T_e \over T_e+\xi T_i } Z^2
        \sqrt{m_e c^2 \over k_B (T_e+\xi T_i)} \Lambda,
\ee
where $\Lambda$ is a Coulomb logarithm, while
$r_0=4\pi r_e^2 m_e c^3$ is a typical energy-loss timescale.

\item To calculate the cooling rates $Q_{cyc}$ and $Cyc_{ei}$
the complete QED crossection of collisional excitation of electron
Landau levels in a strong magnetic field adopted from the work of Langer (1981)
was integrated numerically.
In order to ensure reasonable accuracy in a wide range of temperatures
and magnetic fields, ten Landau levels were taken into account.

\item To calculate $Cyc_{ee}$ the following approximation was adopted from the work
of Langer and Rappoport (1982).
\bea
& Cyc_{ee} = &  2.04 r_0 n_e^2 B_{12}^{-1/2}
                \sqrt{k_B T_e \over \hbar \omega_B}\times \\
&            &  \times \exp\{-{m_e c^2 \over k_B T_e}
                (\sqrt{1+0.04531 B_{12}}-1) \}
                {\Bigl({B_{12} \over 5}\Bigr)}^{{({k_B T_e \over 9597 \,
               {\rm keV}})}^{0.2}}.
\nonumber
\eea

\item To calculate the Bremsstrahlung cooling rate $Br_{ee}$ we
numerically integrated the crossection of Haug (1975) and constructed
the following approximation:
\be
Br_{ee}\approx 2.5410\cdot 10^{-37} T_e^{1.45811} n_e^2 \cdot g(B,T_e),
\ee
where
$g(B,T_e) =
    (0.409-0.0193 B_{12}-0.00244 B_{12}^2){(k_B T_e / 10\,{\rm keV})}^{0.25}$ --
is a gaunt-factor adopted from the work of Langer and Rappoport (1982).

\item To calculate the Bremsstrahlung cooling rate $Br_{ei}$ we
numerically integrated the crossection of Berestetskii and Landau (1982)
for high electron temperatures and adopted the approximation from
the work of Langer and Rappoport for low electron temperatures.
\vspace*{0.1cm}
\be
Br_{ei}\approx \cases{
       0.36\, \alpha \, r_0 (T_e/T_e^b)^{0.5} n_e n_i Z^2 g(B,T_e), &
       when T$_e<$ T$_e^b$  \cr
       0.36\, \alpha \, r_0 (T_e/T_e^b)^{1.2} n_e n_i Z^2 g(B,T_e), &
       when T$_e\geq$ T$_e^b$ }
\ee
\vspace*{0.1cm}
where T$_e^b$=5$\cdot$10$^8$ K.

\item $Q_{relax}$ denotes Coulomb relaxation of the accreting flow
on the electrons of the dense and thin atmosphere of a NS.
The simplest model of an isothermal atmosphere is used and  $Q_{relax}$
is defined as
\be
Q_{relax} = - \nu_{ei} {k_B n_i \over \gamma_i-1} (T_i - T_{atm}),
\ee
where $\nu_{ei}$ is the frequency of Coulomb collisions.

\item $Q_{compt}$ denotes cooling of the flow electrons
in single Compton scatterings. If a scattering photon is not too hard
(${\gamma E_\gamma \over m_e c^2} \ll 1$), the energy lost by an
electron in a single scattering is
$\Delta E=-{E_{\gamma}^2 \over m_e c^2}+{4k_B T_e E_\gamma \over m_e c^2}$.
As $<E_\gamma>=3 k_B T_\gamma,\ <E_{\gamma}^2>=12 k_B T_\gamma^2$,
\bea
& Q_{compt} & = n_e n_{\gamma}<\sigma_T v_{rel} \Delta E> H(B,T_\gamma)= \\
&           & = 12 n_e n_{\gamma}\sigma_T c k_B T_\gamma k_B
                {T_e-T_\gamma \over m_e c^2} H(B,T_\gamma),
\nonumber
\eea
where $H(B,T_\gamma)=(1+0.0165 (\hbar \omega_B / k_B T_\gamma)^{2.48})/
(1+0.0825 (\hbar \omega_B / k_B T_\gamma)^{2.48})$ is the gaunt-factor adopted
from the paper of Arons, Klein, and Lea (1987), while  $n_\gamma$ and $T_\gamma$
are local values of photon temperature and energy density.

\end{enumerate}


\newpage


\end{document}